\documentclass[useAMS,usenatbib]{mn2e}

\usepackage{graphicx}

\title[Clear sky fraction analysis]{Analysis of the fraction of clear sky at La Palma and Mt.Graham
sites}

\author[A. della Valle et al.]{A. della Valle$^{1,4}$, Y. Maruccia$^{2,3}
$\thanks{E-mail:ylenia.maruccia@studio.unibo.it}, S. Ortolani$^{1}$,
 V.Zitelli $^{2}$\\
$^{1}$Department of Astronomy, University of Padova, Vicolo
dell'Osservatorio 3, I-35122, Padova, Italy\\
$^{2}$INAF-Osservatorio Astronomico di Bologna, via Ranzani 1, I-40127, Bologna, Italy\\
$^{3}$Department of Astronomy, University of Bologna, via Ranzani 1,
I-40127, Bologna, Italy\\
$^{4}$Agenzia Regionale per la Prevenzione e la Protezione
Ambientale Veneto, Via Marconi 55, Teolo (PD, Italy)}

\begin{document}

\date{Accepted 2009 Month 00. Received 2009 January 12; in original form 2009 January 12}

\pagerange{\pageref{firstpage}--\pageref{lastpage}} \pubyear{2009}

\maketitle

\label{firstpage}

\begin{abstract}
The amount of available telescope time is one of the most important
requirements when selecting astronomical sites, as it affects the
performance of ground-based telescopes. We present a quantitative
survey of clouds coverage at La Palma and Mt.Graham using both
ground- and satellite-based data. The aim of this work is to derive
clear nights for the satellite infrared channels and to verify the
results using ground-based observations. At La Palma we found a mean
percentage of clear nights of $62.6$ per cent from ground-based data
and $71.9$ per cent from satellite-based data. Taking into account
the fraction of common nights we found a concordance of $80.7$ per
cent of clear nights for ground- and satellite-based data. At
Mt.Graham, we found a $97$ per cent agreement between Columbine
heliograph and night-time observing log. From Columbine heliograph
and the {\it Total Ozone Mapping Spectrometer-Ozone Monitoring
Instrument (TOMS-OMI)} satellite, we found that about $45$ per cent
of nights were clear, while satellite data (GOES, TOMS) are much
more dispersed than those of La Palma. Setting a statistical
threshold, we retried a comparable seasonal trend between heliograph
and satellite.

\end{abstract}


\section{Introduction}
The identification and characterization of a site for the future
European Large Telescope (E-ELT) is a key issue. Moreover a
quantitative survey of cloud cover for the areas selected as
candidate sites for the telescope is and will continue to be an
essential part of the process of site selection for future large
telescopes in the same class as the E-ELT.

In fact, the performance of large telescopes at optical and
infra-red wavelengths is critically dependent on atmospheric cloud
cover. Cloud cover is a key parameter at the time of site selection
and also affects scientific output during the life of the telescope.
For instance, a night-time seasonal trend of fewer clear sky can
reduce regular access to the sky.

Typically it is possible to quantify the presence of clouds at
telescope sites using ground-based observations that provide a real
time knowledge of the atmospheric condition. The fraction of clear
sky can be determined using either instruments (i.e all-sky cameras)
or observer estimates. Long-term records from many ground-based
telescopes, which list the number of nights available for observing,
are now accessible and it is possible to begin a reliable
statistical study. However, this technique alone is not suitable for
identifying future candidate sites where there are no telescopes.

An easy evolution of this analysis is the use of meteorological
satellites that provide measurements of cloud cover and other
critical parameters for site testing covering large areas with
different spatial and temporal resolution. Taking into account
that, most of the meteorological satellites  are equipped with
similar instrumentation, it is not difficult to compare distinct
sites observed by two or more different satellites. Additionally,
since satellite data archives now cover long time periods, it is
possible to have for each site the trend of these parameters in
both long and short time scale. Erasmus and Sarazin (\cite
{erasmus02}) were among the first to demonstrate the successful
application of satellite data for monitoring, comparison and
forecasting evaluations. Erasmus and van Rooyen (\cite
{erasmus06}) quantified cloud cover at La Palma using Meteosat
satellite and validating them using the ground based measurements
taken at Carlsberg Meridian Telescope (CMT).

In this paper we present the results of a study of cloud cover
using satellite- and ground-based data obtained in two different
important astronomical sites: the Observatorio del Roque de Los
Muchachos (ORM) located in La Palma (in Canary Islands), hosting
several international telescopes and among of them the Galileo
National Telescope (TNG), and Mount Graham (in Arizona), hosting
the Large Binocular Telescope (LBT). The results are compared with
Erasmus and van Rooyen (\cite {erasmus02}) and Erasmus and Sarazin
(\cite {erasmus06}). The present paper is organized as follows. In
Section 2, we describe both the ground and satellite data bases
adopted. In Section 3, we describe the satellite data acquisition
procedure, and in Section 4 we show the data reduction procedure.
Section 5 gives a discussion of the results.

\section{DATA}

\begin{figure}
  \centering
  \includegraphics[width=8.5cm]{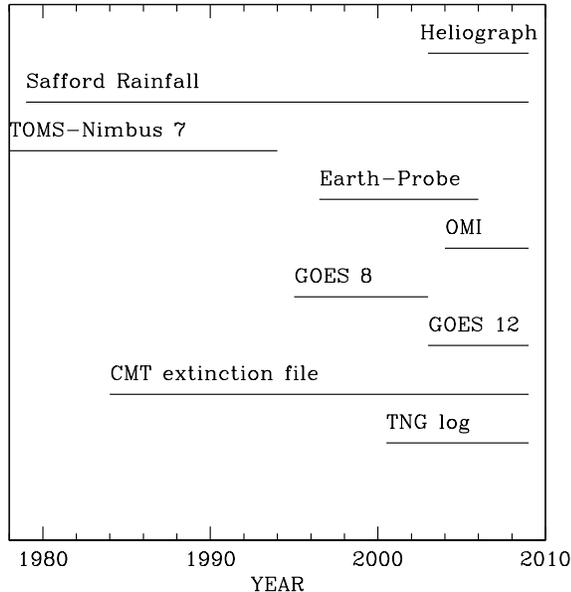}
  \caption{Time coverage of the data bases used.}
             \label{refsatellite}
   \end{figure}

The primary aim of this work is to derive the number of clear
nights. To quantify the amount of clear sky over ORM and Mt.Graham
sites we used different set of data collected at both ground and
satellite facilities available partially via World Wide Web and
partially thanks to the courtesy of the observatory staff. The
validation of satellite data are also performed via correlations
between ground-based and satellite-based data. In particular we used
meteorological satellites that have geostationary orbits because
these ensure large coverage of the globe and a suitable resolution.
Fig.~\ref{refsatellite} shows the time coverage of the data bases
used.

\subsection{Ground-based data}

\begin{table}
 \centering
 \begin{minipage}{170mm}
  \caption{Annual mean downtime as a result of the weather at TNG.}
   \label{ORM-downtime}
  \begin{tabular}{@{}ll@{}}

 \hline
 Year  & Annual mean down time\\
       &  (per cent) \\
\hline
 2000 & 26.7   \\
 2000 &  26.6 \\
 2000 & 26.1  \\
 2000 & 28.2  \\
 2000 & 37.3  \\
 2000 & 39.4  \\
 2000 & 30.2  \\
 2000 & 26.4   \\
 2000 & 29.6   \\
\hline
\end{tabular}
\end{minipage}
\end{table}

The first detailed analysis of more than 10 yr of meteorological
data obtained using the TNG meteorological station at ORM, can be
found in the following two  papers: Lombardi et al.
(\cite{lombardi06}) (hereafter Paper I) and Lombardi et al.
(\cite{lombardi07}) (hereafter Paper II). Paper I shows a complete
analysis of the vertical temperature gradients and their correlation
with the astronomical seeing. In contrast, Paper II shows an
analysis of the correlation between wind and astronomical parameters
as well as the overall long-term weather conditions at ORM.
Differences in the microclimate at the ORM have been demonstrated in
a detailed comparison between synoptic parameters taken at three
different locations at the observatory on a spatial scale of about 1
km. Moreover, the ORM is shown to be almost dominated by high
pressure and characterized by an averaged relative humidity lower
than $50$ per cent. The first detailed reports on night-time
cloudiness at La Palma were given by Murdin (\cite{murdin85}), who
reported $78$ per cent of nights at La Palma were usable during the
period of 1975 February-September (see his Table 2 in Murdin 1985).

Thanks to the kindness of the staff at the TNG and LBT staff, the
updated end-of-night report have been available allowing us to
extend the time baseline to the range 1975-2008 for the TNG
telescope. The LBT logbook, available from National Institute for
Astrophysics (INAF) and Max Planck Institute time, is only available
since 2008 and this gives a time baseline only for two years.

\begin{figure}
  \centering
  \includegraphics[width=7.5cm]{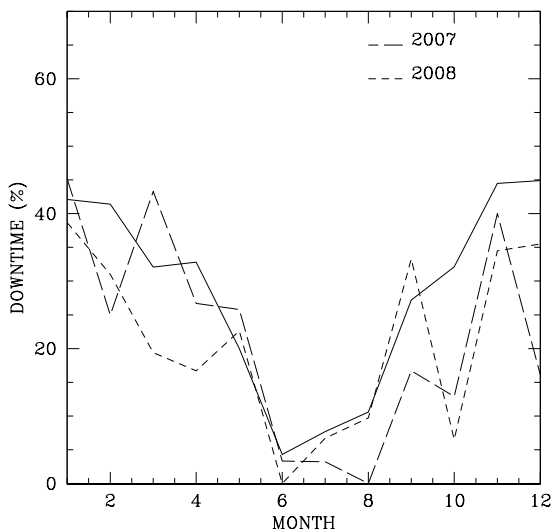}
  \caption{ Distribution of the mean monthly weather downtime at TNG for 2007 (long dashed line)
   and 2008 (short dashed line) from the TNG logbook.
  The continuum line shows the mean downtime computed for 2000-2008.}
             \label{tngdowntime}
   \end{figure}

Fig.~\ref{tngdowntime} shows the distribution of the mean monthly
values for nights lost as a result of bad weather (i.e. cloudy
nights or with strong wind or nights affected by the calima) for
2007 (long dashed line) and 2008 (short dashed line), derived from
the first analysis of the TNG logbook. The continuum line of
Fig.~\ref{tngdowntime} shows the monthly  mean value computed from
2000 to 2008. It is evident that June is the month that has the
minimum number of bad nights. The maximum number of nights lost as a
result of the weather does not reach $50$ per cent of the total
allocate nights. Table \ref{ORM-downtime} reports the annual mean
values of the downtime computed from 2000 to 2008. We can see that
the mean value is almost stable in these last nine years, giving a
value of $30$ per cent of lost nights.

We have also analyzed for the two years  2007 and 2008 all the
astronomical useful nights at TNG using two different criteria. In
the first criterion we extracted the information from the {\it lost
time weather} string reported in the end-of-night report. If the
night was fully used, no lost time should be found in this string.
Using this information, we separated the nights used into fully
useful nights (i.e. the dome is open the whole night) and partially
useful nights (i.e. less than 5 h are lost because of the weather).
The results are given in Table \ref{ORM-nights}. As discussed in the
following, the adopted criterion is more stringent than that of
Arderberg (\cite{arder83}).

\begin{table*}
 \centering
 \begin{minipage}{100mm}
  \caption{Mean monthly percentage of the useful nights at TNG.
  Selection is derived using the number of observed hours extracted from the TNG log. }
   \label{ORM-nights}
  \begin{tabular}{@{}lcccc@{}}
  \hline
           &\multicolumn{2}{c}{2007} & \multicolumn{2}{c}{2008}\\
           &100 per cent used& Used $>$ 5 h. &100 per cent used& Used$>$5h\\

 \hline
 January   &25.8 &29.0 & 35.5 &29.0   \\
 February  &46.4 &24.1 & 20.7 &24.1  \\
 March     &38.7 &12.9 & 32.2 &22.6   \\
 April     &46.6 &26.7 & 40.0 &23.3  \\
 May       &70.9 &9.7  & 54.8 &29.0   \\
 June      &73.3 &10.0 & 86.7 &13.3    \\
 July      &90.3 &6.4  & 87.1 &9.7   \\
 August    &90.3 &3.2  & 67.7 &22.6    \\
 September &73.3 &16.7 & 33.3 &20.0  \\
 October   &38.7 &32.2 & 35.5 &22.7  \\
 November  &20.0 &16.7 & 48.3 &20.7   \\
 December  &51.6 &16.1 & 32.2 &19.3   \\
\hline
 mean      &55.5 &17.0 & 47.8 &21.3 \\
 \hline

\end{tabular}
\end{minipage}
\end{table*}

From Table \ref{ORM-nights} it can be seen that the TNG telescope is
opened for the whole night and without interruptions for more than
$50$ per cent of all the nights considered. If we compute the mean
percentage of the two years, considering both totally and partially
useful nights, we find that the telescope operated the $70.8$ per
cent of the total nights. We can conclude that this value is in
agreement with the values expected for good astronomical sites.
Erasmus and van Rooyen (\cite {erasmus06}) gave a percentage of 74.7
per cent for usable nights at Cerro Tololo Inter-Americn Observatoy
for the period 1997 June to 1999 April, derived from ground-based
observations. Moreover, at the ORM,  Arderberg (\cite{arder83})
measured $47$ per cent of nights to be photometric in 1982 and $67$
per cent in 1983. In his analysis, Arderberg defined as {\it
photometric} every night having at least six hours of uninterrupted
clear sky. The mean value of $57$ per cent of photometric nights,
computed in 1982 and 1983 by Arderberg (\cite{arder83}) is in
agreement with our mean value of $51.7$ per cent, obtained using the
more restricted criterion of $100$ per cent fully used nights. This
means that in case of six consecutive hours of clear sky, a full
night is very likely to be photometric according to Arderberg's
definition.

In the second criterion, we classified the nights using the {\it sky
condition} comments. We classified the nights as clear (i.e.
cloud-free) and mixed (i.e. partially used because of the presence
of clouds during the night). We also took into account the calima.
In this type of selection, if the night presents strong wind or
humidity, it appears mixed in our classification, while it may
appear not usable using the previous criterion because the dome may
be closed for safety. Table \ref{ORM} shows the mean monthly
percentage obtained. A comparison between Table \ref{ORM-nights} and
Table \ref{ORM} shows that the percentage of fully used nights is
similar to the percentage of $100$ per cent clear nights. This means
that when a night is fully used it is very likely to be completely
clear. However, table \ref{ORM} gives a higher number of total
usable nights. This is probably  because, as we have just said, the
mixed sky conditions may include also strong wind or humidity
conditions. We notice that a $14.5$ per cent of nights in 2008 were
lost as a result of high humidity or strong wind.

The same classification has been done for LBT telescope. Table
\ref{LBT} shows the distribution obtained for the quality of nights.
The LBT logbook is limited to a sample of only 50 nights (July and
August do not have data because the telescope is closed due to the
monsoon), but for the completeness of the discussion we decided to
collect them in Table \ref{LBT}. A comparison between Table
\ref{ORM} and Table \ref{LBT} shows a comparable percentage of clear
days, while at Mt.Graham the percentage of mixed nights is higher.
Unlikely to La Palma site, no calima events has been found at
Mt.Graham.

 The criterion to discriminate between photometric and
spectroscopic is not unique. It is difficult, for example, to judge
at the beginning of the night before starting the observations
whether the sky is completely clear, in sense of clouds free, or
whether the airborne dust will significantly affect the observing
conditions. For this specific point, we used the CMT extinction
files to set the clear sky quality at La Palma site.

Regarding Mt.Graham, an interesting criterion was used, based
instead on the morning sky conditions (Steward Observatory,
\cite{arizona87}), but is not used in this analysis. Thanks to the
weather station of the United States Forest Service (USFS) located
2.1 km far from the LBT (Columbine peak), we correlated the log of
night observations with data from the heliograph an instrument that
records the Sun's radiation in the wavelengths from UV to IR. This
data base covers all the period since 2001 September.

We also used the rainfall data base of the weather station of
Safford Agriculture Center\footnote {See
http://www.wrcc.dri.edu/cgi-bin/cliMAIN.pl?azsaff} because of its
longtime baseline, since 1940. The cross check gives complementary
results to those of Columbine peak.

\begin{table}
 \centering
 \begin{minipage}{80mm}
  \caption{ Mean monthly percentage of useful nights at TNG. The selection
  is derived  using the sky quality comments extracted from the TNG logbook.
  The fraction is relative to the total number of the available nights per month.}
   \label{ORM}
  \begin{tabular}{@{}lllllll@{}}
  \hline
   Name     &  \multicolumn{3}{c}{year 2007}& \multicolumn {3}{c}{ year 2008} \\
   Month & Clear &Mixed &Calima& Clear & Mixed& Calima  \\
         &  (\%) & (\%) & (\%) & (\%) & (\%) & (\%) \\
 \hline
 January   &25.8 & 25.8 &3.2 &19.3 &35.5 & 6.4 \\
 February  &53.6 & 17.9 &3.6 &17.2 &41.4 & 10.3\\
 March     &26.7 & 20.0 &10.0& 35.5 &29.0& 16.1\\
 April     &56.7 & 13.3 &3.3 & 46.7 &30.0& 6.7 \\
 May       &58.1 & 6.4  &9.7 & 48.4 &29.0& 0   \\
 June      &83.3 & 13.3 &0   & 83.3 &6.7 & 10.0\\
 July      &61.3 & 0    &35.5& 67.7 &0   & 25.8\\
 August    &93.5 & 3.2  &3.2 & 45.2 &9.7 & 35.5\\
 September &76.7 & 6.7  &0   & 33.3 &33.3& 0   \\
 October   &51.6 & 35.5 &0   & 38.7 &54.8& 0   \\
 November  &16.7 & 43.3 &0   & 48.3 &17.2& 0   \\
 December  &58.1 & 25.8 &0   & 33.3 &32.3& 0   \\
\hline
 mean      &55.2 & 17.6 &5.7 & 43.1 &26.6& 9.2 \\
 \hline
\end{tabular}
\end{minipage}
\end{table}

\begin{table}
 \centering
 \begin{minipage}{80mm}
  \caption{Monthly sky conditions at Mt.Graham in 2008 using the sky quality
  comments of the LBT logbook. The monsoon months are excluded.}
   \label{LBT}
  \begin{tabular}{@{}llll@{}}
  \hline
       &  \multicolumn{3}{c}{ 2008} \\
    & Clear & Cloudy & Mixed  \\
    &  (per cent) & (per cent) & (per cent)  \\
\hline
  Mean     & 60.0&10.0 & 30.0 \\
\hline
\end{tabular}
\end{minipage}
\end{table}

\subsection{Satellite-based data}

\begin{table*}
 \centering
 \begin{minipage}{120mm}
  \caption{The mean parameters of the used satellites.}
   \label{tabsatellite}
  \begin{tabular}{@{}lp{0.8cm}p{0.5cm}cp{1.2cm}p{1cm}l@{}}
  \hline
   Name     &  Long.& Lat. &  range & Spatial  &Temporal  &  used years \\
            &       &       & $\mu m$   &Res. [km]     &Res. [h] & \\
 \hline
 GOES 12     & -75$^{\circ}$& 0$^{\circ}$& 6.5 - 7.0 (B3)   & 8 &3 & 2007-2008 \\
              &     &    & 10.2 - 11.2 (B4) & 4 &3 & 2007-2008\\
 GOES 8       & -122$^{\circ}$&   -5$^{\circ}$  & 6.5 - 7.0 (B3)   & 8 &3 & 2007-2008 \\
              &     &    &  10.2 - 11.2 (B4) & 4 &3 & 2007-2008\\
 Nimbus7      & polar    &     & 0.32 - 0.40   & 111        & 24 & 1978-1993\\
 Earth-Probe  & polar    &    & 0.32 - 0.40   &111   &24   & 1996-2005\\
 OMI          & polar    &    & 0.35 - 0.50   & 3 &24 &  2004 - present   \\
 \hline
\end{tabular}
\end{minipage}
\end{table*}

Data were derived using {\it Geostationary Operational Environmental
Satellite (GOES) 12} for ORM and GOES 8 and GOES 12 for Mt.Graham,
which are meteorological satellites monitoring cloud cover and water
vapour. Both satellites are the new generation of GOES family, an
American geosynchronous weather facilities of the National Oceanic
and Atmospheric Administration (NOAA). GOES 12 is able to observe
the full Earth disk in both visible and IR regions of the
electromagnetic spectrum and can observe and measure cloud cover, in
addition to other important meteorological parameters. Two GOES
satellites are typically used to provide coverage of the entire
hemisphere. When the two satellites are in operation, one satellite
covers the GOES east position, located over the Equator at
75$^{\circ}$E, and the other one is located at GOES west position
over the Equator at 135$^{\circ}$W. These two satellites provide
imagery of North and South American continents as well as Pacific
and Atlantic Oceans with an overlapping area of coverage. GOES 8 was
launched in 1994 April and operated from 1994 November to 2003April.
GOES 12 was launched on 2001 July and replaced GOES 8 in 2003 April.
The two spacecrafts carry an imager, a 'sounder' and a X-ray imager.
The imager is a Cassegrain telescope covering five wavelength
channels, one in the visible bands ($0.55-0.75~ \mu m$), and four in
the infrared ($3.80-4.00$, $6.50-7.00$, $10.20-11.20$, and
$11.50-12.50~\mu m$) bands. It can provide images covering $3,000
\times 3,000$ km$^2$ every 41 seconds, by scanning the area in 16
square kilometer sections. Full Earth-disk scans are scheduled every
3 h. It should be noticed that GOES 12 covers La Palma area near the
edge of the field of view. Taking into account the curvature of the
Earth we obtain a projection of about 57$^{\circ}$ in latitude and
28$^{\circ}$ in Longitude, corresponding to a factor 1.1 in Latitude
and 1.8 in Longitude.

Thanks to this set up, it is possible to have the same instrumental
configuration for the ORM and Mt.Graham and to compare them in a
suitable way.

We have also included data from polar satellite of the {\it Total
Ozone Mapping Spectrometer (TOMS)} family to extend the statistics
of clear days at Mt.Graham. These satellites were planned to study
atmospheric ozone, but their data can be used for cloud cover. Four
TOMS satellites were launched from 1978 to the present. Because of
the failure of the second satellite, {\it Meteor-3}, we have only
used data from the remaining three: {\it Nimbus-7}(1978-1993), {\it
Earth-Probe} (1996-2005), and {\it ozone Monitoring Instrument
(OMI)}; (from 2004). We used the overpass data\footnote{See
ftp://toms.gsfc.nasa.gov/pub/omi/data/overpass/} covering a square
of 110 km$^2$ centered on Tucson city (Arizona). Table
\ref{tabsatellite} summarize the main parameters of the satellites
used.

\section{Satellite data acquisition}

For this work, we used GOES 12 equipped with the imager. Among the
five available channels, we selected the water vapour channel
(channel 3, hereafter called b3 band) centered at $6.7~\mu m$, and
the cloud coverage channel (channel 4, hereafter called b4 band)
centered at $10.7~\mu m$. The b3 band is sensitive between
$6.5-7.0~\mu m$ and is able to detect high-altitude cirrus clouds.
The b4 band is sensitive between $10.2-11.2~\mu m$ and is able to
detect middle-level clouds. The output of the detector is
proportional to the energy reaching the detector areas per unit time
(radiance). It is also possible, given the intensity and the
wavelength of the radiance, to derive the equivalent brightness
temperature using an appropriate Planck function. This was the
procedure adopted by Erasmus and van Rooyen (\cite{erasmus06}).

We selected the IR channel because water vapour absorbs
electromagnetic radiation and then re-emits it in various wavelength
bands, in particular in the infrared region at $6-7~\mu m$. If
clouds are not present, the emissions at $10.7~\mu m$ reaching the
satellite are largely not absorbed  by the atmosphere, so that
radiance values measured are a result of emission from the surface.
However, when clouds are present they behave as absorbing and
emitting "surfaces" so that, under these conditions, radiation
reaching the satellite is from the cloud top, which has a lower
emissivity as a result of a lower temperature.

The GOES data have the highest spatial resolution
($4~\rm{km}\times4~\rm{km}$ for channel b4 and
$8~\rm{km}\times8~\rm{km}$ for channel b3) compared with the old
generation satellites. Data are prepared by the Comprehensive Large
Array-data Stewardship System (CLASS), an electronic library of NOAA
environmental data\footnote{See www.class.ngdc.noaa.gov}, and are
stored as rectified full earth disk images in a format called AREA
files. We processed them using McIDAS-V Version 1.0beta1, a free
visualization and data  analysis software package. The first step to
collect data was to extract the right sector centered close to the
TNG and LBT.

Regarding the TNG, we know that it has latitude $28^\circ45'28.3"$~N
and longitude $17^\circ53'37.9"$~W. We identified and extracted from
the sector a subimage of $1^\circ \times 1^\circ$ having the central
pixel centered on (or near) the TNG coordinates. We obtained
sub-images having the central pixel of $28^\circ 46'32.16"$~N
latitude and $17^\circ 52'0.84"$~W longitude. Both the TNG and
subimage coordinates reside in the same pixel that is also the
instrumental resolution. For each night we took into account
available observations at four different hours  to cover the entire
observed night, at 20:45, 23:45, 02:45 and 05:45 UTC. The mean of
these four values has been used in our analysis.

Regarding Mt.Graham, data from 1995 January to 2008 December have
been analysed: until 2003 April, the data come from GOES 8, and then
GOES 12 has been used. For each night we selected data at two
different hours: at 17.45 and 02.45 UTC (10:45 and 19:45 of local
time, respectively). When no data was available at those time, we
used images taken at different times, up to 1.5 h after or before.
Our aim was to select two hourly sets of data, one during daytime
and one during night-time. However, it was not possible to built
homogeneous series at 05:45  or 08:45 UTC (22:45 and 01:45 of local
time, respectively) because  during spring and autumn the GOES
change acquisition timetable and do not often cover these times.

The analogous data base has been extracted for LBT, located at
$32^\circ 42'33.2"$~N latitude, and $109^\circ 54'7.6"$~W longitude.
We selected subimages with the pixel centered at this position.
After a detailed analysis of the GOES daily signal coincident with
the log data, we found the best correlation with the signal at 10:45
and 19:45 local time (see also Fig.\ref{LBTseason}).

\section{Data analysis}

\begin{figure}
  \centering
  \includegraphics[width=7.5cm]{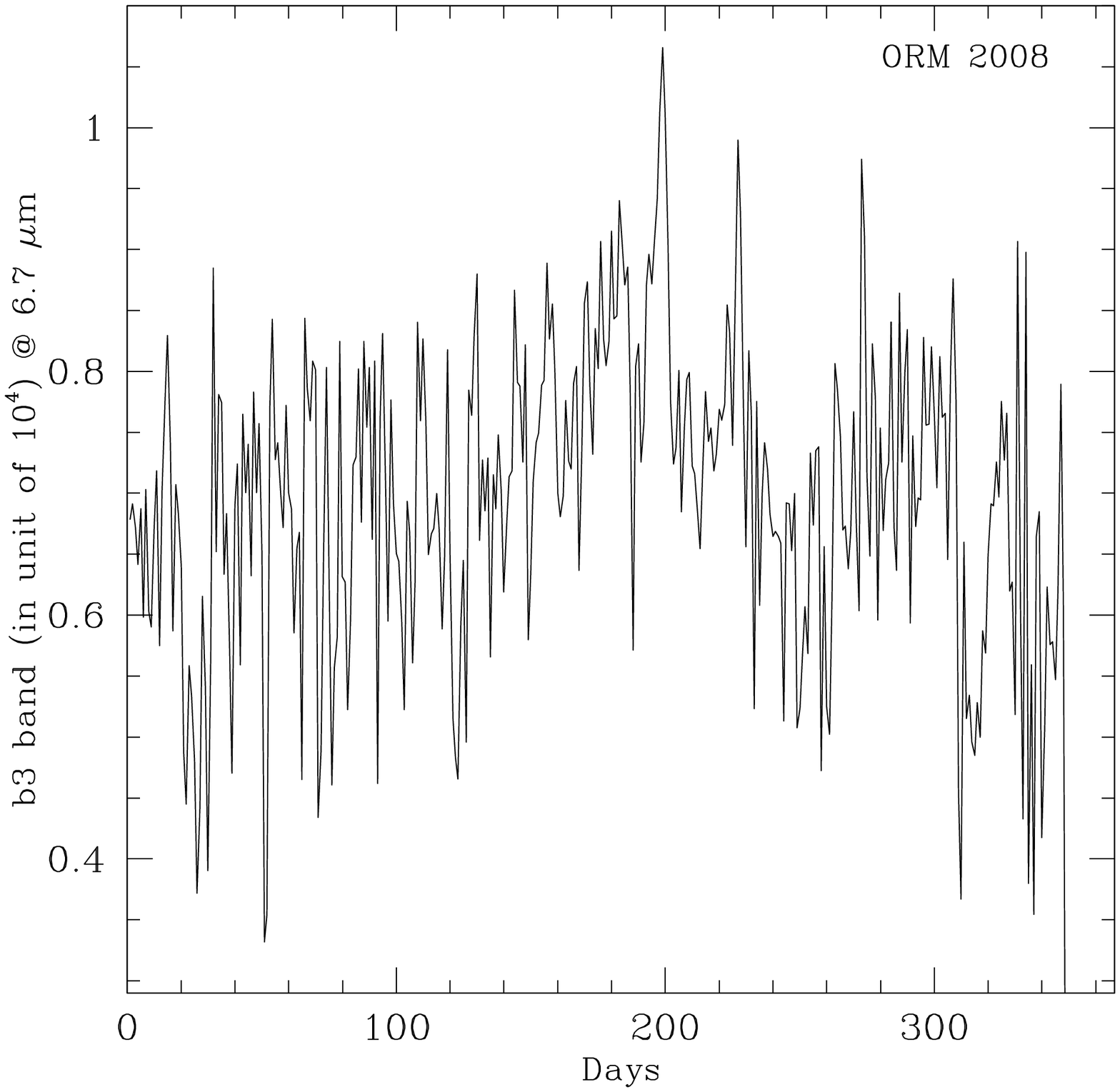}
\includegraphics[width=7.5cm]{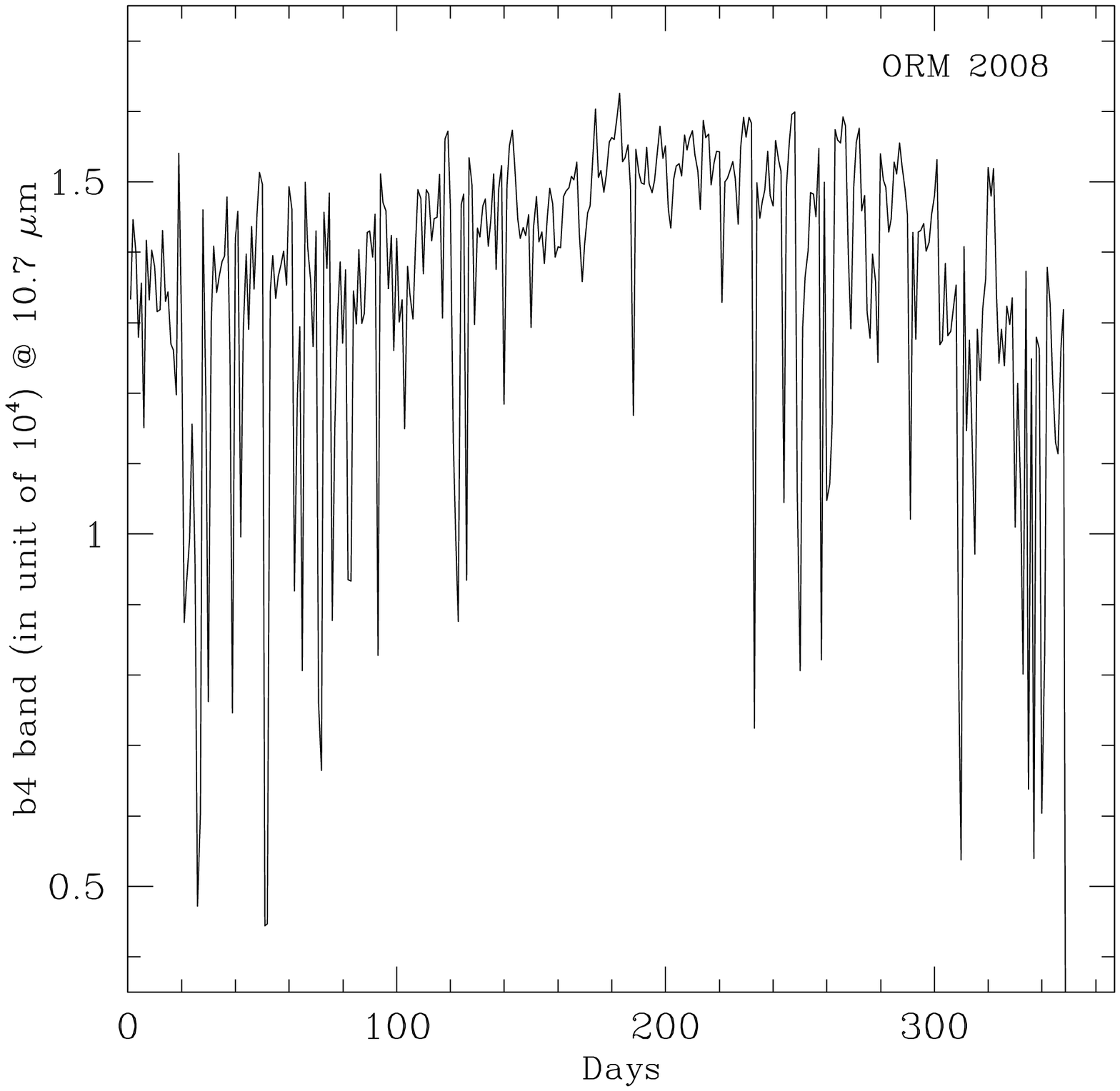}
  \caption{GOES 12 emissivity distribution of the two selected bands at the ORM in 2008. }
             \label{mtgyear2008}
   \end{figure}

Fig.~\ref{mtgyear2008} shows the distribution of the mean infrared
emission at the ORM for the b3 and b4 bands in the upper and lower
panels, respectively, for 2008. It is evident that b3 band ($6.7~\mu
m$ water vapour) shows higher values of emissivity in the summer
time period, corresponding to a higher temperature and a low
percentage of clouds ($180-200$ d), than those in the autumn. The
lower panel of Fig.~\ref{mtgyear2008} for the $10.7~\mu m$ band
shows a flatter distribution of emissivity.

 \begin{figure}
  \centering
  \includegraphics[width=7.4cm]{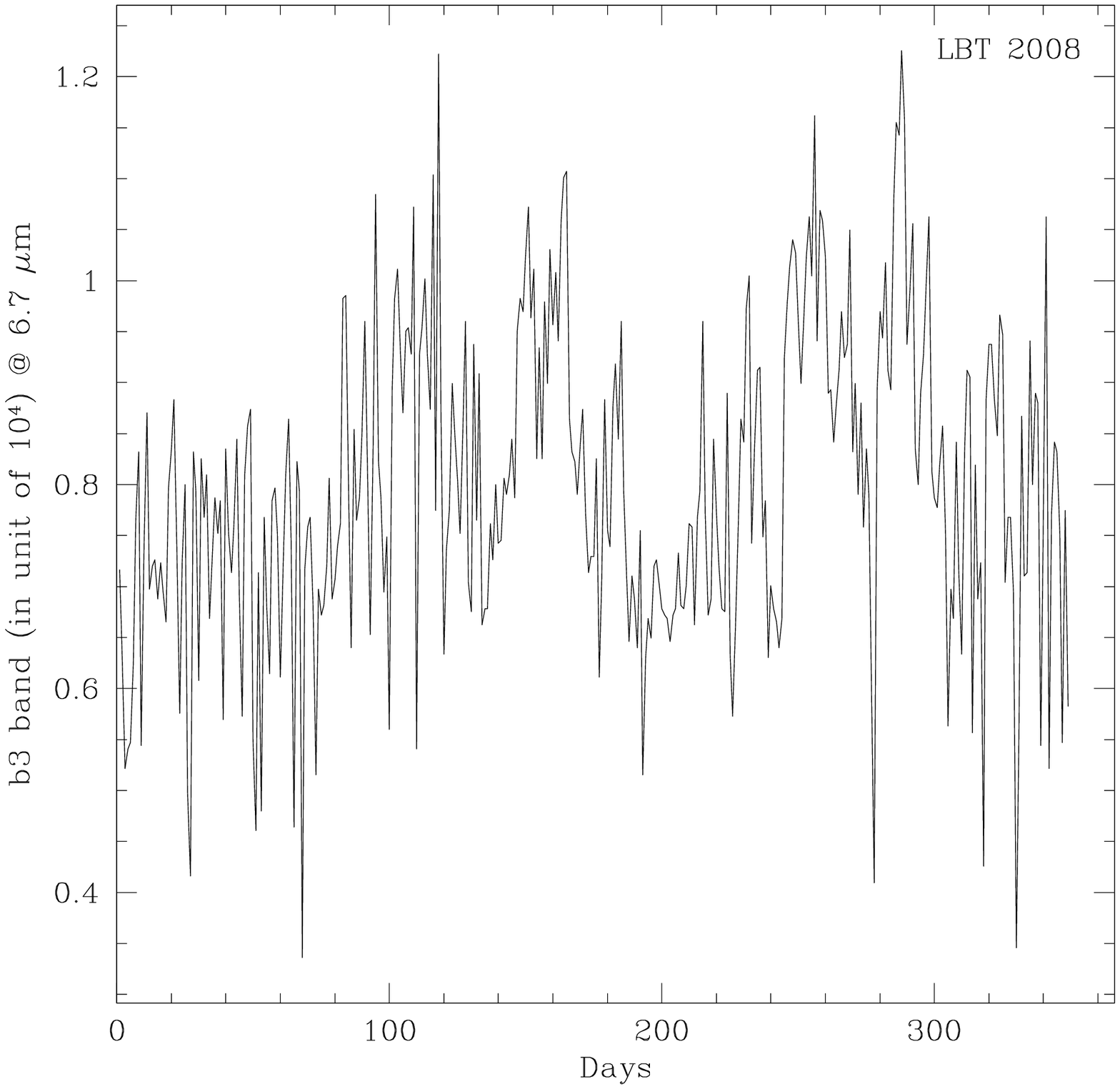}
\includegraphics[width=7.4cm]{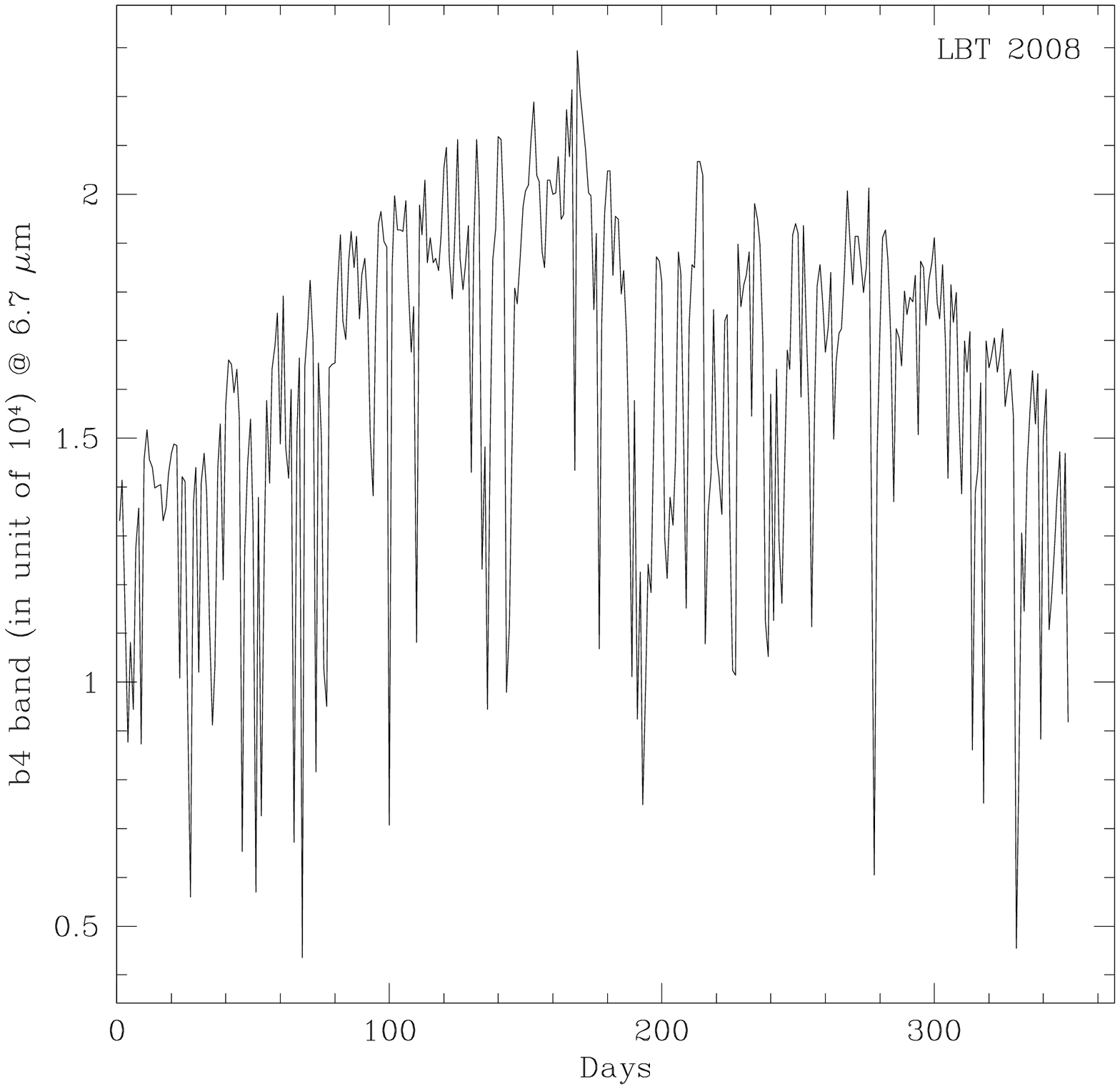}
  \caption{ GOES 12 emissivity distribution of the two selected bands at
Mt.Graham in  2008.}
             \label{lbtyear2008}
   \end{figure}

Fig.~\ref{lbtyear2008} shows the daily distribution of the
emissivity for the  b3 and b4 bands at Mt.Graham. The seasonal
effects of this site are evident in both the bands, distinguishing
the monsoon period. This seasonal trend in b4 band is much more
evident than at ORM. This trend requires the normalization of the
flux in order to allow a selection of the night quality from a
predetermined fixed threshold (see later for discussion). The
monsoon period is seen in July and August. A spline interpolation
was adopted because of the discontinuity of the monsoon period.

\begin{figure}
  \centering
  \includegraphics[width=8.5cm]{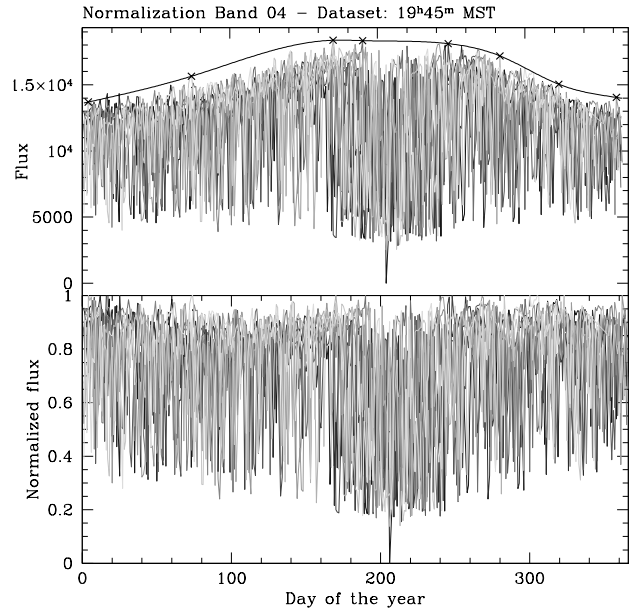}
  \caption{GOES 12 emissivity distribution of the b4 band at Mt.Graham in 2008 of raw data
  (upper panel) and normalized data (lower panel).  }
             \label{paragoes}
   \end{figure}

Fig.~\ref{paragoes} shows the distribution of the emissivity in b4
band ( top) and the distribution of normalized emissivity ( bottom)
centered at Mt.Graham. The normalization function is used to extract
the other values of emissivity under the assumption that the
behavior is the same all year.

\begin{figure}
  \centering
  \includegraphics[width=6.5cm,angle=90]{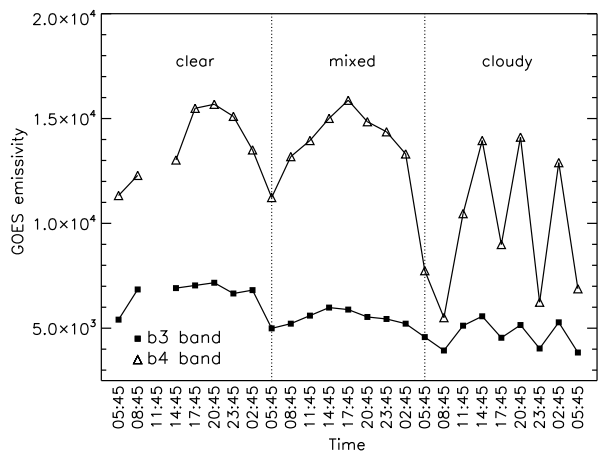}
  \includegraphics[width=6.5cm,angle=90]{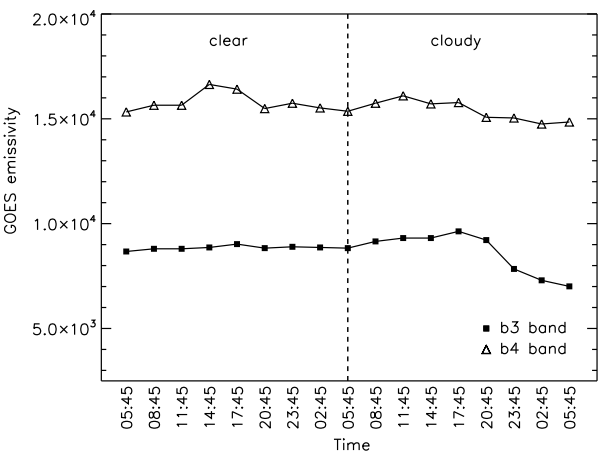}
  \caption{Distribution of GOES 12 b3 band (filled squares) and b4
  band (open triangles) values at the ORM in two different periods of the year.
   The upper panel shows three consecutive days in winter, and the lower panel shows two consecutive days
   in summer.}
             \label{ORMseason}
   \end{figure}

A comparison of the two Fig.~\ref{mtgyear2008} and
Fig.~\ref{lbtyear2008} shows that  the satellite reaches higher
values of emissivity in both bands at Mt.Graham compared to La
Palma. A deeper analysis has to be carried out to check the
existence of correlations with other parameters  to understood this
different value of emissivity between the two sites. In this paper,
we have checked possible seasonal effects, and Fig.~\ref{ORMseason}
shows the distribution of the emissivity at ORM for the b3 (filled
squares) and b4 (open triangles) bands in two different periods of
the year. The upper panel of Fig.~\ref{ORMseason} shows the
distribution of three consecutive days (clear, mixed and cloudy) in
winter time. The clear day reaches a value of about 14000 units in
b4 band and a mean value of about 9700 units in the cloudy day. The
presence of cold cirrus produces the oscillation of the counts in
cloudy day. A similar behavior can be seen in b3 band with a lower
value of counts. The bottom panel of Fig.~\ref{ORMseason} shows the
distribution of two consecutive days in summer time (only clear and
cloudy because we found no mixed and consecutive day). We see that
the clear day reaches a value of about 16000 units in b4 band a
greater value with respect the winter value, probably because of the
different mean air temperature. A more evident effect of the
arriving perturbation can be seen in b3 band with a drop of the
counts.

\begin{figure}
 \centering
  \includegraphics[width=6.5cm,angle=-90]{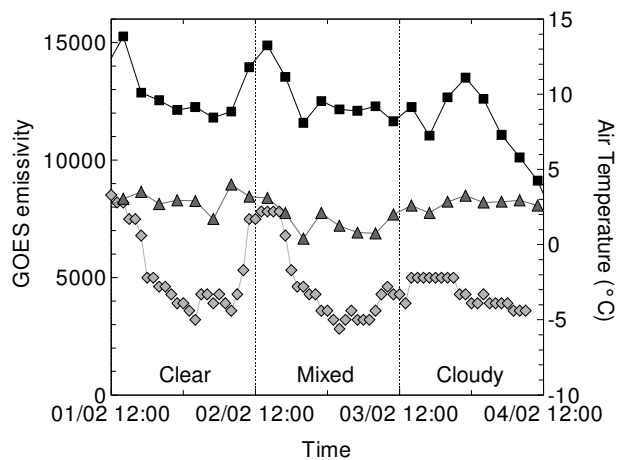}
  \includegraphics[width=6.5cm,angle=-90]{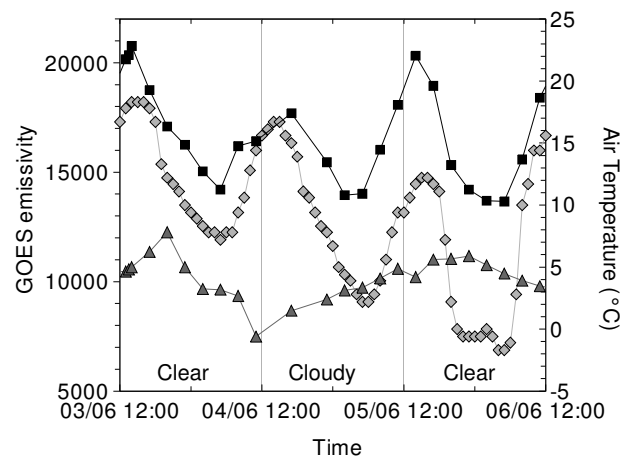}
  \caption{ Distribution of GOES 12 b3 (triangles) and b4 (squares) band values at Mt.Graham during some
  winter (upper panel) and summer (lower panel) days in 2008, compared with the temperature of the air measured
   by Columbine weather station (diamonds, right axis). The astronomer's sky condition string is shown at the bottom of
   the plot (see the text for details).}
             \label{LBTseason}
   \end{figure}

To meke an easy comparison of the seasonal behavior of the two
sites, we include here the analogous distribution for Mt.Graham.
Fig.~\ref{LBTseason}  shows the distribution of b3 (triangles) and
b4 (squares) bands of GOES 12 satellite during some winter (upper
panel) and summer (lower panel) days in 2008, compared with the
temperature of the air measured by the Columbine weather station
(diamonds, right axis). We confirm that the signal from La Palma is
systematically lower than Mt.Graham, in spite of the lower altitude.
A tentative explanation could be a higher extinction of the
satellite signal because of the longer of optical path.

Fig.~\ref{LBTseason} clearly shows the difficulty in finding the
right threshold of GOES fluxes. When the day (and night) is clear,
air temperature and  b4 band values follow a day/night cycle. When a
perturbation arrives, the daily peak usually disappears. However,
the winter graphic shows a peak of b4 band flux during the night,
even if the air temperature remains stable. The satellite probably
measured the temperature of the clouds,which in that case had a
greater temperature than the ground. Using the threshold method to
distinguish clear days from cloudy days, this day would be
identified as clear. In our sample, there is only marginal evidence
(if any) of lower minimum temperatures during covered nights
compared to clear nights. The most evident difference, instead, is a
lower emissivity in the late afternoon just before a cloudy night.
The local, ground temperature, may have an important role on GOES
measurements. Furthermore the higher signal during the summer does
not seem to lead to better discrimination between clear and cloudy
nights. Noise in the data and resolution effects have been tested
also comparing the single pixel data to nine-pixels averages during
nights recorded from the ground; however, the plots are almost
identical. More tests are needed to check the reliability of nights
time GOES at Mt.Graham.

\subsection{La Palma: the threshold method}
Considering the sky quality comments as derived from the
end-of-night report at TNG, we correlate the values of b3 and b4
bands of GOES with the corresponding night quality.
Fig.~\ref{reflapalma} shows the distribution of the emissivity
obtained from  b3 (x-axis) and b4 (y-axis) bands as a function of
the different type of nights for 2007 (filled triangles) and 2008
(open squares). Each type of night is plotted separately to better
identify the distribution along the panels. {\ Fig.~\ref{reflapalma}
shows good correlation between TNG classification and emissivity. In
fact, it is evident that at higher b3 and b4 band values it
generally corresponds to a clear night, and at lower b3 and b4 band
values it corresponds to a cloudy night. In contrast, mixed nights
have a distribution among all the possible values, probably due to
the season.

 \begin{figure}
  \centering
  \includegraphics[width=7.5cm]{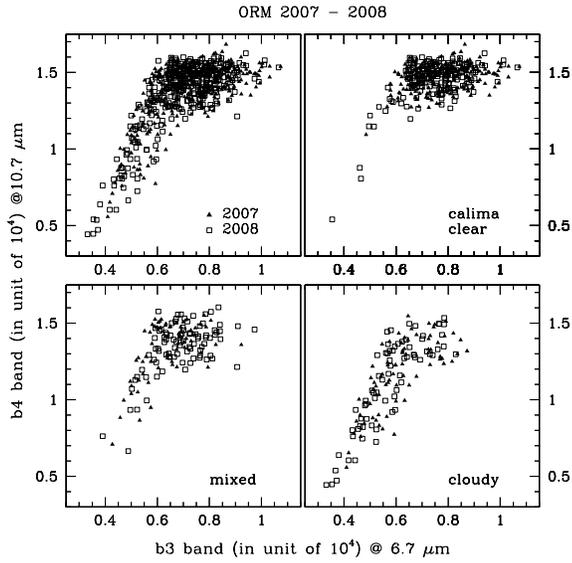}
  \caption{ Distribution of GOES 12 emissivity at ORM separated according to the different
   sky quality (extracted from the TNG log). The years 2007 (filled triangles) and 2008 (open squares) are plotted.}
             \label{reflapalma}
   \end{figure}

In order to have a more objective analysis of clear conditions, we
used and cross-correlated the same satellite data with the
corresponding atmospheric extinction as published in the website of
the CAMC telescope\footnote {See
http://www.ast.cam.ac.uk/$\sim$dwe/SRF/camc\_extinction.html}. In
this analysis, we presume that sky conditions ( with the two
telescope having a close locations) were the same and extinction
parameter is able to identify not only useful nights but also the
photometric nights. The nightly values of extinction were derived
from CCD frames in the Sloan Digital Sky Survey (SDSS) r' band.

Each frame contains an average of $30-40$ photometric standard
stars. In our analysis, we assumed that if no value of extinction
existed or if it was equal to zero, it was probable that no
observation took place because of the bad sky conditions. Nights
with technical problems have not been included (233 nights).

 Fig.~\ref{extyear2008} shows the distribution of the extinction as
a function of GOES b3 (upper panel) and b4 (lower panel) band
emissivity. The points represent all the nights with an extinction
value not equal to zero and are classified by extracting information
from TNG logbook for 2007 and 2008. They show a large spread of the
extinction values in both infrared bands, some of which have been
classified as cloudy at TNG. Clear nights are located in a
well-defined locus of the  b4 band. It is interesting to note that
$72.1$ per cent of all the nights reporting calima on the TNG report
in 2007 and 2008, have extinction values greater then 0.13.
Moreover, those nights having reported no calima but some clouds
have extinction values greater or equal to 0.13 but lower b3 and b4
values than the nights with calima. Most of our selected clear
nights (88 per cent) present an astronomical extinction less than
0.13 mag. The extinction value of 0.2 mag airmass$^{-1}$ on clear
nights was found to be discriminant for calima events in a previous
paper of Lombardi et al. (\cite{lombardi08}).

 \begin{figure}
  \centering
  \includegraphics[width=8.5cm]{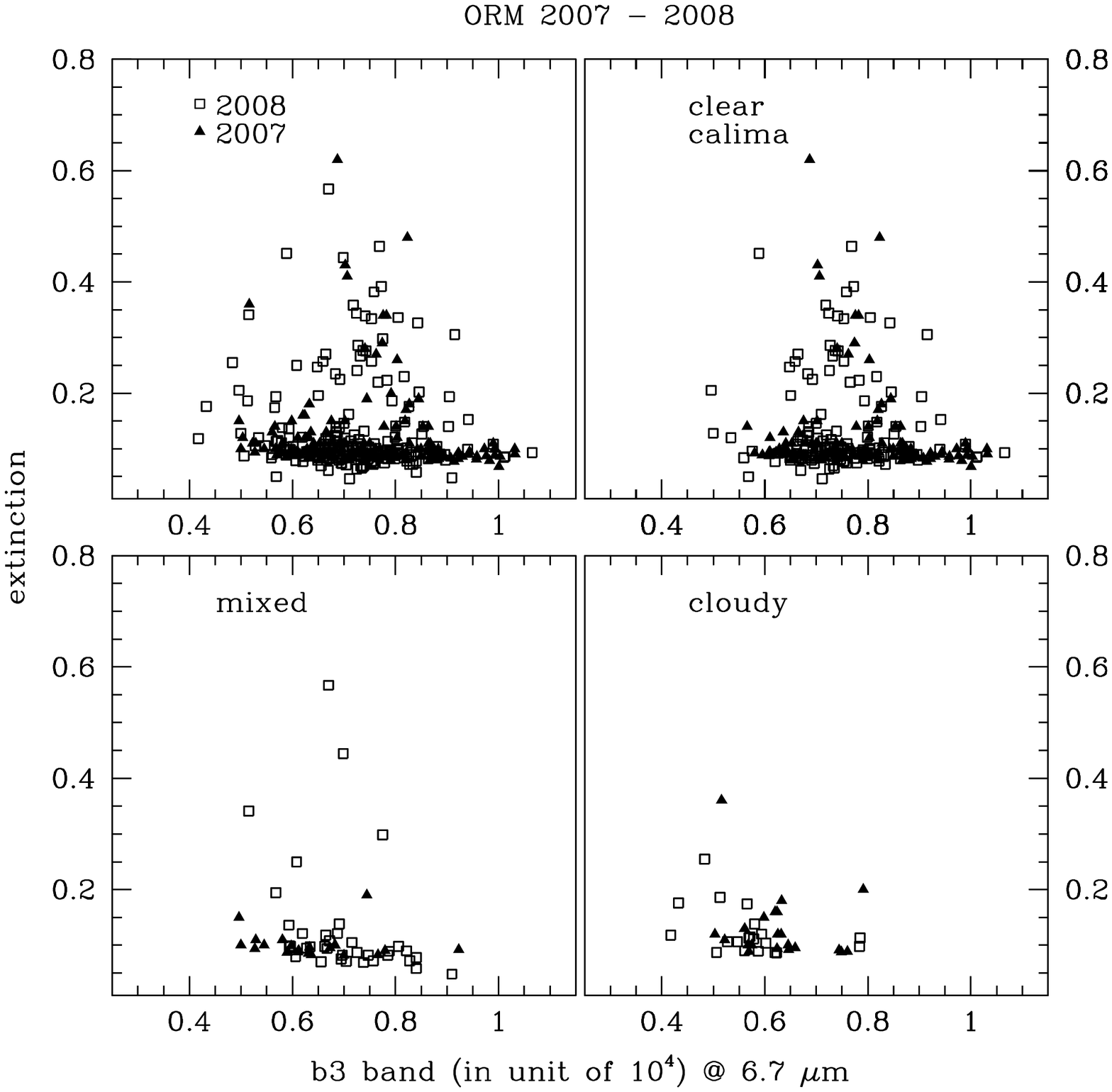}
\includegraphics[width=8.5cm]{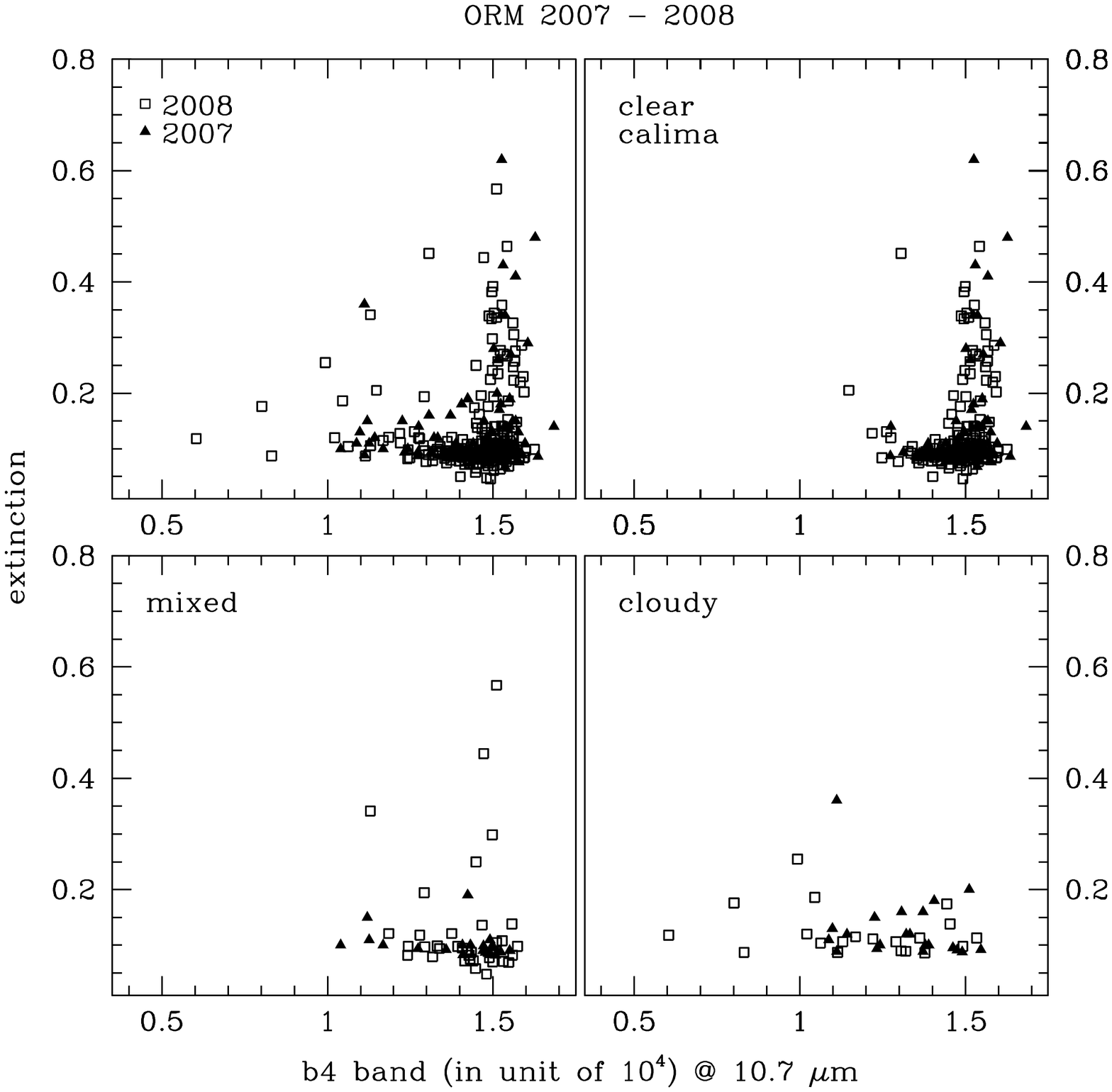}
  \caption{ Distribution of the extinction as a function of GOES 12 b3 band (upper panel) and b4 band (lower panel)
  emissivity only for nights with extinction in 2007 and 2008. Classification has been carried out using the TNG log.}
             \label{extyear2008}
   \end{figure}

Fig.~\ref{distrgettati08} shows the distribution of the GOES 12
emissivity  as a function of TNG night report only for those nights
not plotted in Fig.~\ref{extyear2008} because no extinction value
was reported (about 123 nights with "zero extinction"
classification). We can see that the majority of the plotted points
are located in the panels marked mixed and cloudy. In fact,
considering the TNG logbook for 2007 and 2008, we found that the
$63.4$ per cent of these nights with no extinction at CMT are
classified as cloudy or mixed ones, while only the $36.4$ per cent
are classified as clear nights or with the calima.

\begin{figure}
  \centering
  \includegraphics[width=8.5cm]{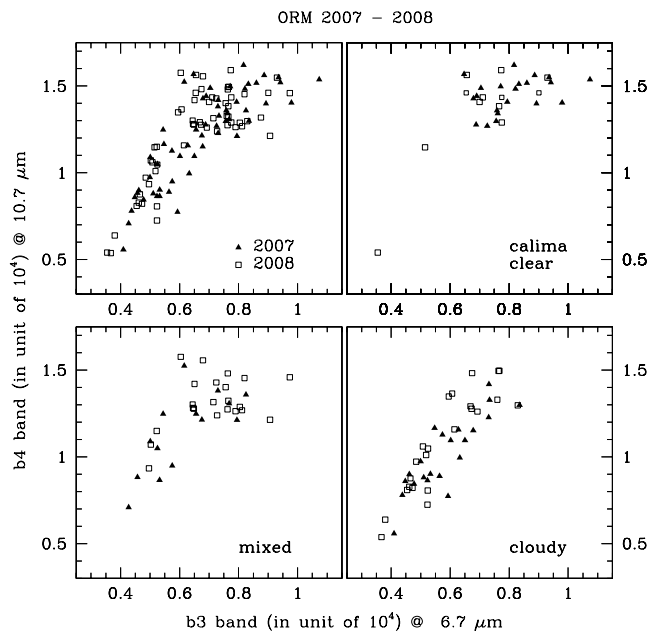}
  \caption{ Emissivity distribution of GOES 12 b3 and b4 band emissivity at the ORM in 2007 and 2008
   for all the nights with "zero extinction"  classification in the CMT log (no observations).
   Sky quality classification has been carried out using the TNG log.}
             \label{distrgettati08}
   \end{figure}

To conclude we decided to use Fig.~\ref{extyear2008}  to define the
threshold values for both b3 and b4 band to distinguish satellite
clear nights from cloudy ones.  In particular we decided to obtain
the satellite sample of clear nights, setting the GOES threshold  to
6500 for b3 band and 13200 for b4 band. The ground sample is
obtained using all the nights with CMT extinction values. The result
is shown in Fig.~\ref{refcompleto} where dashed lines indicate the
threshold that we choose to separate clear nights. All GOES nights
presents in b3 band values greater than 6500 and in b4 band values
greater than 13200 are clear, all the nights with b3 band values
less than 6500 and b4 band values less than 13200 are cloudy, and
the other cases are mixed. This choice has been adopted by
optimizing the discrimination of the different nights and minimizing
the contamination. It should be noticed that this empirical
criterium is different from the method used by Erasmus based,
instead, on the derived temperatures. With this adopted limits, we
obtained the following statistics: GOES identified $73.6$ per cent
of clear nights for 2008 and $70.2$ per cent for 2007, while from
ground observation we know that the $61.6$ per cent of all nights
were clear for 2008 and the $63.6$ per cent for 2007. Thus, it seems
that, using our GOES processing method, clear nights are
overestimated of about $10$ per cent.

\begin{figure}
  \centering
  \includegraphics[width=8.5cm]{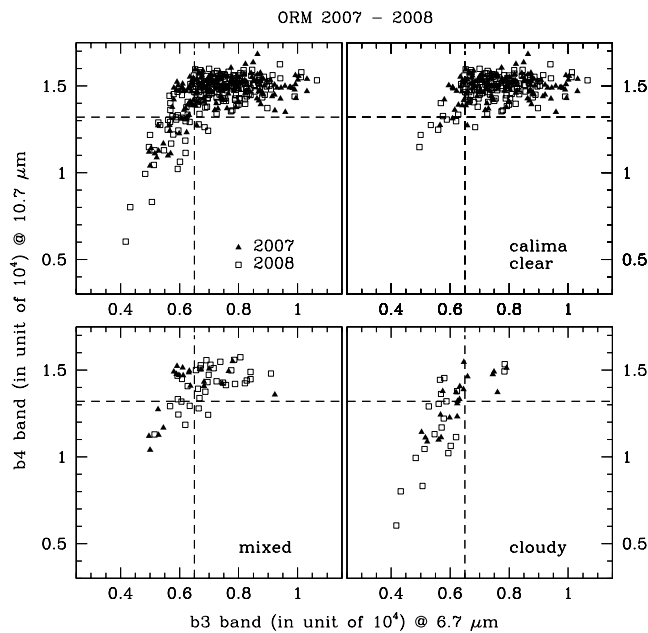}
  \caption{ Distribution of GOES 12 b3 and b4 band emissivity  at La Palma in  2007 and 2008
  for all nights reporting an extinction value different from zero in CMT extinction file.
  Sky quality classification has been carried out using the TNG log. The dashed lines indicate the thresholds chosen to
   separate clear nights from mixed and cloudy nights.}
             \label{refcompleto}
   \end{figure}

\begin{figure}
  \centering
  \includegraphics[width=8.5cm]{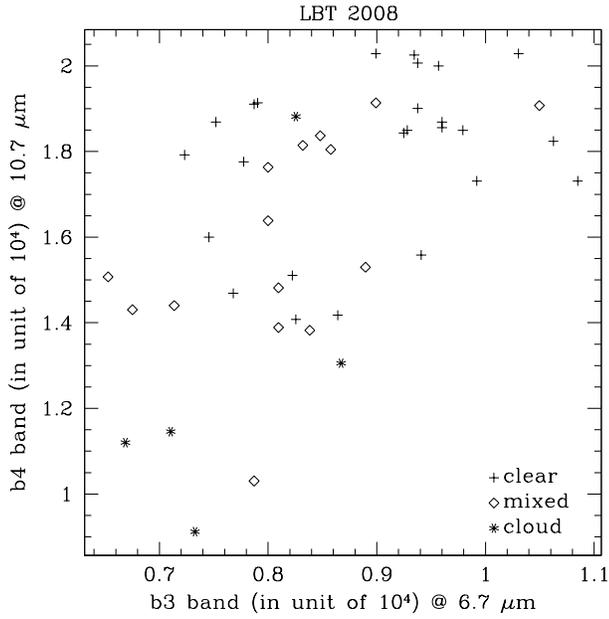}
  \caption{ Distribution of GOES 12 b3 and b4 band emissivity at Mt.Graham.
  Sky quality classification has been carried out using the LBT report.}
             \label{lbtemission}
   \end{figure}

\subsection{Mt.Graham: a new approach}

Fig.~\ref{lbtemission} shows the results of the same adopted
procedure that correlate nights quality based on LBT report and GOES
emissivity. The separation between clear, mixed and covered nights
on the base of b3 and b4 GOES bands is not as evident as la Palma
data. This could in part be the result of an intrinsic GOES data
interpretation at Mt.Graham, and in part a result of poor
statistics. In order to improve the latter, after a number of tests,
we decided to follow a new different approach using the heliograph
data from Columbine peak. The archive is maintained by the Western
Regional Climate Center\footnote{ see the web page
http://www.wrcc.dri.edu/cgi-bin/rawMAIN.pl?azACOL} and data are
freely downloaded. For the purpose of this paper, we downloaded the
daily table from 2001 March  (from when the heliograph data are
available). This database is not complete: for some days no data
were recorded or were only partially recorded; for the first case
the event is highlighted, but not the second. So we verified the
reliability of the data checking day per day the data base and
classifying days for completeness: {\it perfects} if data covered
the entire day, {\it good} if only one hour data lacked, {\it bad}
if more than one hour was not recorded. We found that perfect days
make up $64.2$ per cent of the total, good $8.7$ per cent, and bad
days $27.1$ per cent. We only used perfect and good days in this
analysis, covering $72.9$ per cent of the considered period.
Fig.~\ref{helio} shows the distribution of Sun emissivity, as
integrated daily fluxes, for 2008, where the bad days appear as flux
drops.

 \begin{figure}
  \centering
  \includegraphics[width=8.5cm]{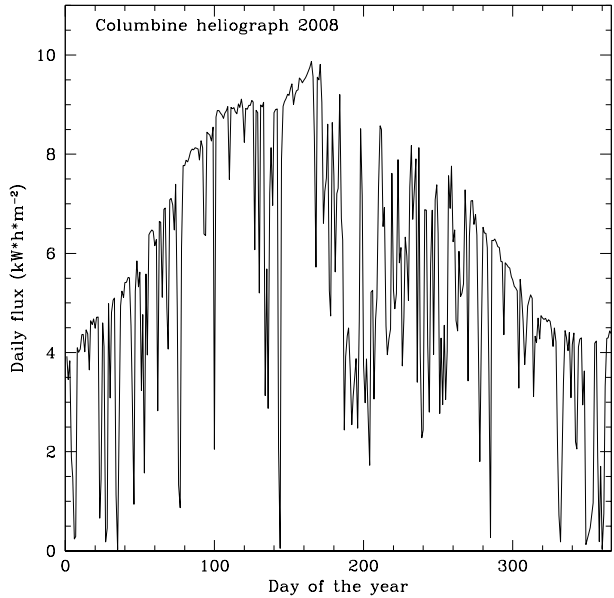}
  \caption{ Yearly distribution of Sun emissivity, from integrated daily fluxes, for 2008.
  The bad days appear as flux drops.}
             \label{helio}
   \end{figure}

The monsoon period (day 180 to about 240) is evident. Because of the
strong seasonal effect, the heliograph data have been normalized
like the GOES data using a spline fit. Finally the normalized daily
fluxes have been compared with the end-of-night reports of LBT. The
correlation is presented in Fig.~\ref{helio_split} where the
correspondence of the heliograph flux with the night-time data is
very tight, with a concordance of $97$ per cent of clear nights when
the threshold of the normalized flux is set at $0.9$. It is evident
that the  day/night difference at Mt.Graham is almost negligible if
we consider only clear days, but not when considering  mixed/cloudy
days, where $30$  per cent of the samples fall in the other
category.

 \begin{figure}
  \centering
  \includegraphics[width=8.5cm]{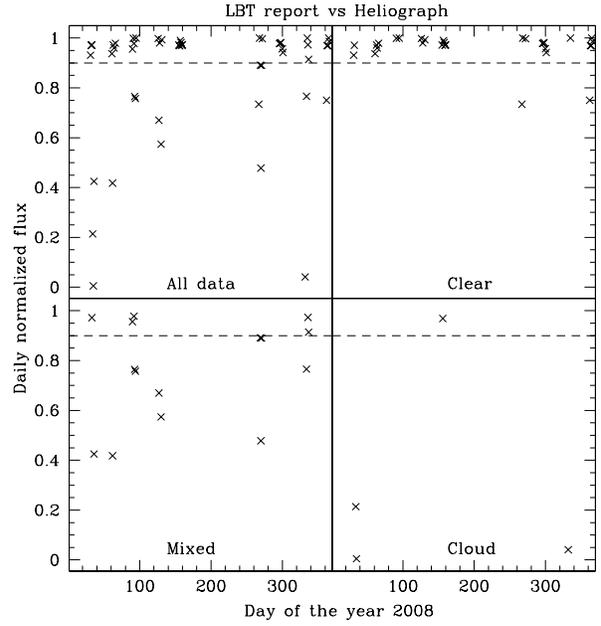}
  \caption{ Daily heliograph flux distribution at Mt.Graham separated according to the
  different sky quality (extracted from the LBT report. Dashed lines indicate the thresholds
  chosen to separate the clear nights from the mixed and cloudy nights.}
             \label{helio_split}
   \end{figure}

 \begin{figure}
  \centering
  \includegraphics[width=8.5cm]{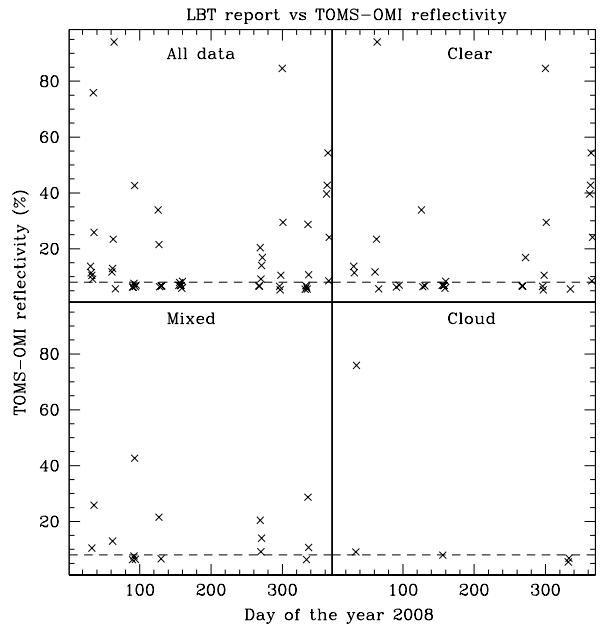}
  \caption{Daily TOMS-OMI reflectivity distribution at Mt.Graham separated according to the
  different sky quality (extracted from the LBT report).}
             \label{refl_split}
  \end{figure}

A similar procedure was used to compare the TOMS-OMI reflectivity
with the logbook of LBT. In this case the best threshold was found
to be 0.08. The results are visible in  Fig.~\ref{refl_split}.
 It is evident that heliograph and
TOMS-OMI data distinguish clear nights better than GOES satellites.
The upper-right panels of Figs.~\ref{helio_split} and
~\ref{refl_split} show, respectively, the flux and the reflectivity
distribution of the clear nights. Mixed nights (lower-left panels),
instead, are less separable. Dashed lines indicate the threshold
that we choose to separate clear from mixed and bad nights.

Because the observation logbook is available only for 2008 and its
concordance with the heliograph is very good, we decided to use the
heliograph data to extend the reference data in the past, to 2002,
while TOMS data are available from 2004. Then we compared both GOES
b3 and b4 normalized fluxes with the heliograph data. We found two
different thresholds, where the percentage of clear nights found by
GOES was comparable with that obtained with TOMS-OMI and the
heliograph in the overlapped period. The two values are 0.66 and 0.8
of the b3 and b4 normalized flux respectively for the heliograph and
for the TOMS.

\section{Discussion}
\subsection{La Palma site}
In our analysis at La Palma site we used 731 nights obtaining the
sky comments from TNG end-of-night reports. After the
cross-correlation of the TNG nights with the CMT extinction, the
total number of nights used for the final statistics was 375: these
we can define photometric or, at least, spectroscopic on the basis
of Table \ref{ORM-nights}. We obtained 700 nights from GOES
database, covering the period 2007-2008. Fig.~\ref{refcompleto}
shows the  thresholds adopted to select night quality. All those
having emissivity greater than 6500 counts in b3 band and a values
greater than 13200 counts in b4 band are defined as clear nights. In
contrast, we defined as cloudy all the nights with b3 band values
less than 6500 and b4 band values less than 13200; all other cases
are classified as mixed. For ground observations, we defined clear
nights all those ones with no clouds, humidity, strong wind or
calima.

\begin{table}
 \centering
 \begin{minipage}{80mm}
  \caption{ Mean monthly percentage of clear nights selected from both ground and
  satellite for 2007 and 2008. The selection is computed using the logbook
  for the ground and the established threshold values for satellite data. The monthly fractions are computed only
  for the  nights with extinction values.}
   \label{tabfinale}
  \begin{tabular}{@{}lllll@{}}
  \hline
   Name  &  \multicolumn{2}{c}{ 2007}& \multicolumn {2}{c}{  2008} \\
   month & CMT    & GOES & CMT   & GOES\\
         &  (per cent)  & (per cent) & (per cent)  &  (per cent)  \\
 \hline
 January   &43.8 & 43.8 &  -  &  -  \\
 February  &65.2 & 60.9 &  -  &  -  \\
 March     &30.0 & 60.0 &66.7 &100.0\\
 April     &  -  &   -  &59.1 &68.2 \\
 May       &33.3 & 66.7 &53.8 &73.1 \\
 June      &100.0& 100.0&89.5 &100.0\\
 July      &58.3 & 100.0&75.0 &96.4 \\
 August    &91.3 & 100.0&50.0 &92.3 \\
 September &81.5 &63.0  &62.5 &87.5 \\
 October   &57.7 &57.7  &55.0 &85.0 \\
 November  &75.0 &50.0  &66.7 &33.3 \\
 December  &  -  &  -   &38.1 & 0.0 \\
\hline
 mean      &63.6 & 70.2 &61.6 &73.6 \\
 \hline
\end{tabular}
\end{minipage}
\end{table}

With these adopted limits  we found a mean statistical percentage of
clear nights of $62.6$ per cent from ground data and $71.9$ per cent
for satellite data. It seems that there is a quite good correlation
between ground and satellite data, but the satellite overestimates
the clear nights by about $10$ per cent, as already found by Erasmus
and van Rooyen (\cite {erasmus06}).

The next step was to quantify the nights classified as clear using
both ground and satellite data. We found that the $79.7$ per cent of
the common nights (365, i.e. half of the total sample) are clear for
both ground and satellite data for 2007, and $81.6$ per cent for
2008, with a mean percentage of $80.7$ per cent. Table
\ref{tabfinale} shows the distribution of the monthly percentage of
the common clear nights at La Palma in the years 2007 and 2008.
Moreover we note that most of the clear nights have an extinction
less than 0.13. Months with $"-"$ indicate that no data are
available from GOES database or from CMT extinction file. There is a
large spread of the agreement during the months and also in the two
considered years.

Using the Meteosat in geostationary orbit at $0^\circ$, at the ORM
Erasmus and van Rooyen (\cite {erasmus06}) found 68.7 per cent of
clear nights (i.e. cloud free) from ground based data, on the basis
of 629 nights in the period of 1999-2002, compared with our
percentage of $62.6$ per cent. In contrast, they found a percentage
of $65.0$ per cent from satellite data, compared with our mean of
$71.9$ per cent. Moreover, they found 52.5 per cent of photometric
nights for both satellite and ground data, a much lower value than
our mean value of $80.7$ per cent. We note that Erasmus and van
Rooyen (\cite {erasmus06}) report 83.7 and 85.3 per cent of
photometric hours  for satellite and ground data, including the time
when only part of the sky was photometric. A summary of the number
of photometric nights can be found at
www.otri.iac.es/sitesting/UserFiles/File/photometric-time.pdf .

Instead, the estimated percentage of spectroscopic nights is
different. If we assume that the partially used nights are
spectroscopic, we found from the logbook a mean value of about
$16.8$ per cent. Erasmus and van Rooyen found a mean value of $20.7$
per cent. Instead, from satellite they found that $23.8$ per cent
are spectroscopic nights, compared with our result of about $14.5$
per cent. Finally, we found that the $31.0$ per cent are
spectroscopic nights for both the satellite and the ground-based
data, compared with $8.3$ per cent given by Erasmus and van Rooyen.

Finally, we considered the nights affected by the calima. We found
that about $80$ per cent of nights affected by the calima are
considered clear for GOES, the satellite does not distinguish the
calima.

A test has been carried out to verify if the satellite and ground
data are more in agreement when considering nights affected by the
calima and the clear nights together, rather than considering only
clear nights. Under this assumption, we obtained that the clear
nights (i.e. cloudless nights) plus nights affected by the calima,
are $68.8$ per cent for ground selection for 2008 (compared with
$73.6$ per cent for GOES) and $73.2$ per cent for 2007 (compared
with $70.2$ per cent for GOES). The mean percentage is $71.0$ per
cent, showing a very good agreement between satellite data and
ground-based observation. We can conclude that satellite does not
distinguish the calima event. Fig.~\ref{lapalmayear} shows this
result, where the monthly distribution of the clear night fraction
is plotted from TNG and CMT logbooks (dashed line) and GOES 12
(solid line). The agreement of the two plots is excellent except for
December.

The final step was to understand how many nights were clear for both
ground and satellite data in this database including nights affected
by the calima. We found that the $81.9$ per cent of all the clear
nights are effectively clear for both GOES and ground for 2007 and
the $80.3$ per cent for 2008.  We found a mean percentage of $81.1$
per cent, compared with the previous mean of $80.7$ per cent of
clear nights not including the calima. It is evident that the
percentages are similar and this is probably because of the low
number of nights affected by the calima. We emphasize that the
satellite selects clear nights without making assumptions about the
transparency of the sky.

 \begin{figure}
  \centering
  \includegraphics[width=6.5cm,angle=90]{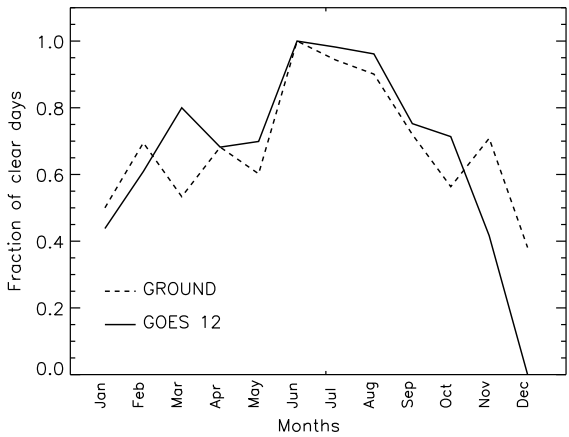}
  \caption{ Composite  distribution of clear nights, even if affected by the calima,
  at La Palma from both ground (dashed line, from TNG and CMT logbook) and satellite (solid line, GOES 12) data. The
  fractions are computed for the subsample of CMT nights with extinction values.  }
             \label{lapalmayear}
   \end{figure}

\subsection{Mt Graham site}

At Mt Graham the total number of clear days analyzed in the period
2002-2008 are 912 from the heliograph and 964 from GOES. The common
number of clear days is 661, with a relative fraction of $72.0$ per
cent of common days versus the total number of clear days from the
heliograph. Fig.~\ref{LBTmonth} shows the monthly composite
distribution of the monthly average fraction of clear night computed
from the heliograph database, the Toms-Omi satellite, GOES data
taken at 17:45 and 02:45 UTC (10:45 and 19:45 local time,
respectively), data from Mt Graham site testing (Steward Observatory
\cite{arizona87}) and the rain distribution derived from Safford
(Agriculture center) database. With the obvious exception of Mt
Graham site testing, all the data are from the period of 2007-2008.
There is a general qualitative agreement among the different methods
(the rainfall is clearly anticorrelated with the clear night
fraction) with the heliograph data. A high peak in May is evident,
as well the sharp cut-off between May and June, and a secondary peak
in autumn. This trend is in very good agreement with the Kitt Peak
logbooks of photometric time fraction analyzed by D.L.Crawford
(Steward Observatory \cite{arizona87}; see also the data of usable
time available at http://www.noao.edu/kpno/usrhnbk/user-App.html).
The GOES data are higher than the average of the other data in
September and October and lower in May. While the fraction of usable
time ($44.5$ per cent) is in agreement with the other methods, the
distribution is clearly more noisy.  The 1982-1983 site testing data
are in a very good agreement with the heliograph data, except during
October-December where these data are much lower. On average the
site testing data are lowest. We recall that winter time in Arizona
presents high variability, connected with the episodic invasions of
storms coming from north-west. The resulting yearly usable fraction
is between $43.0$ per cent and $46.0$ per cent. It should be noticed
that GOES data analysis from Erasmus \cite{erasmus02} at Mt Graham
gives a usable fraction much higher that all our indicators ($61.0$
per cent clear and $60.0$ per cent usable), in particular during the
monsoon season. This suggests that GOES data interpretation at Mt
Graham presents some unresolved multiparametric problems and the
results are very sensitive to the data treatment. The difficulty in
the quantitative interpretation of GOES data is extensively
discussed in literature (Stove et al. \cite{stowe}, Khyier et al.
\cite{meteo04}, and Jung et al. \cite{jung}).

\begin{figure}
  \centering
  \includegraphics[width=8.5cm]{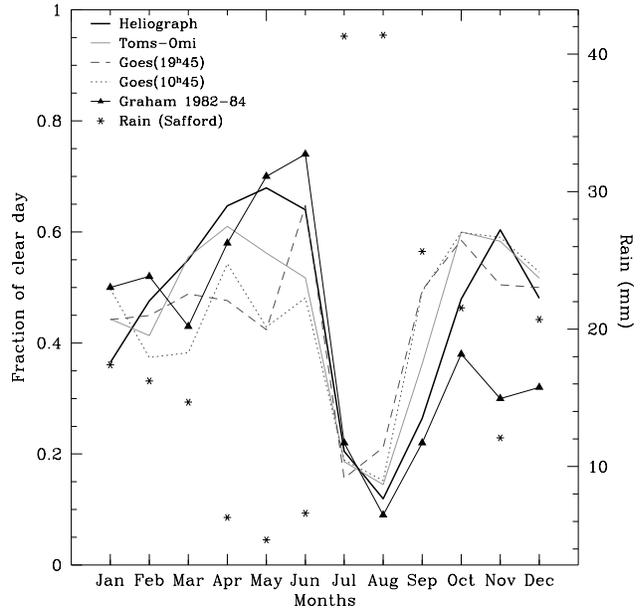}
  \caption{Composite monthly mean distribution of clear nights at Mt Graham }
             \label{LBTmonth}
   \end{figure}

 \begin{figure}
  \centering
  \includegraphics[width=8.5cm]{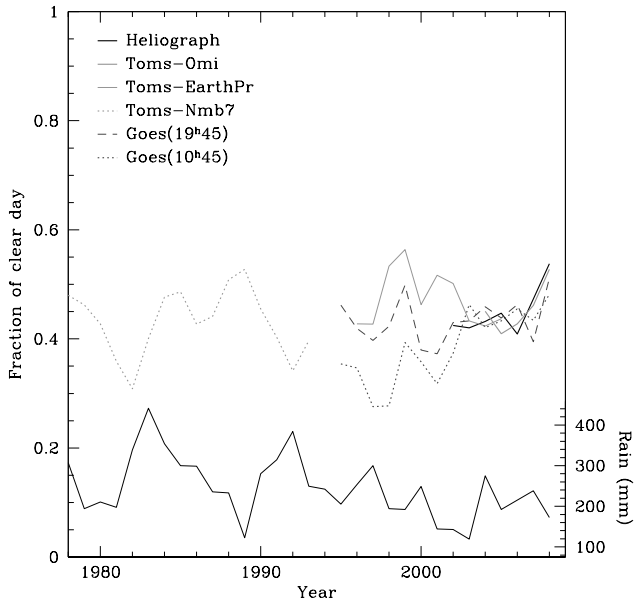}
  \caption{Composite yearly mean distribution of clear nights at Mt Graham.
  The lower plot shows the rainfall yearly distribution at Safford. }
             \label{LBTyear}
   \end{figure}

Fig.~\ref{LBTyear} shows the long-term yearly composite distribution
of the fraction of clear night fraction computed using the
heliograph database, the {\it TOMS (OMI, Earth probe and Nimbus7)}
satellite data, the GOES 8 and 12 data taken at the 19:45 and 10:45
local time and the rain distribution derived from Safford database.
It can be seen as the rain distribution is very well anticorrelated
with the distribution of the clear time, in particular with the
clear time record in the database of {\it Nimbus-7}. A major drop in
the fraction of clear time occurred between 1982 and 1983,
corresponding to a sharp rainfall peak. This was also the period of
the Mt Graham site testing. According to the suggestion discussed in
the site testing report we confirm that the lower fraction reported
(and discussed above) was due to a climate fluctuation. A general
trend with decreasing rainfall and increase of the clear time
fraction between 1980 and 2008 can also be seen. The year 2008 (in
particular the spring) was recorded in Arizona as one of the driest
of the last fifty years.

\section{Conclusion}

We have presented a quantitative survey of cloudy coverage at la
Palma and Mt Graham using both ground and satellite data. In order
to quantify the amount of clear nights, we used different databases,
in particular the end-of-nights reports obtained at TNG and LBT
telescopes, the CMT extinction file for La Palma and the heliograph
and rainfall databases at Mt Graham. Satellite data are derived from
GOES 12 at La Palma and GOES 8 and 12 at Mt Graham. A further check
have been done only for Mt Graham using the TOMS family satellites.

The analysis mainly addresses the years 2007 and 2008 but a
long-term analysis is also reported for Mt Graham. The sample at La
Palma is composed of 731 consecutive nights. After the cross
correlation with the extinction file, this is reduced to 365 nights.
The analysis at Mt Graham is based on 912 days. The sample from
satellite is composed of 700 nights at la Palma and 964 at Mt
Graham. A fixed threshold in the GOES IR emissivity selects the
clear nights by satellite. At La Palma we have obtained that $62.6$
per cent of the 365 sampled nights are selected clear from ground,
and $71.9$ per cent of the 365 nights are selected clear from
satellite. Taking into account the common nights between ground and
satellite data we found that $81.1$ per cent of the nights are
selected clear for both. This shows good agreement but indicates
that about $19$ per cent of the clear nights from ground are lost by
the satellite data. At Mt Graham we found a good agreement between
Columbine heliograph and night time observing log. In this case,
that satellite found only $72.0$ per cent of the total of clear days
found by the heliograph.

Two relevant additional conclusions can be derived from the Mt
Graham analysis, as follows:
\begin{itemize}
\item \textbf the rainfall trend at Safford can be used as tracer of the night time
conditions at Mt Graham and possibly also in the whole Arizona area.
\item \textbf the limited day/night weather evolution at Mt Graham makes the results of
a very simple device as the heliograph, useful to monitor with high
accuracy the local status of the night time clear sky.
\end{itemize}
In addition, at la Palma we can derive the following conclusions:
\begin{itemize}
\item \textbf It is possible to define a threshold in satellite emissivity to
select clear nights with an uncertainty of $20$ per cent.
\item \textbf A good correlation exists between GOES 12 satellite and ground-based data.
\item \textbf Using the common selected nights we found that $80.7$ per cent are classified
clear from both GOES 12 satellite and ground logbook.
\item \textbf The marginal increase to $81.1$ per cent of the concordance obtained
including calima events confirm that the satellite is able to
distinguish only the presence of clouds.
\end{itemize}
A further analysis of the satellite data (e.g.a wider field, the
simultaneous use of different channels, etc.) is suggest to improve
the prediction of clear nights. Furthermore, we are studying the
fraction of clear nights lost for high humidity or strong wind.

\subsection{ACKNOWLEDGMENTS}
The authors acknowledge dr.Vincenzo Testa, from National Institute
for Astrophysics, Roland Gredel, from Heidelberg Max Planck
Institute for Astronomy, Antonio Magazzu from TNG and the former TNG
Director Ernesto Oliva for the collaboration.

This activity is supported by the European Community (Framework
Programme 7, Preparing for the construction of the European
Extremely Large Telescope, Grant Agreement Number 211257).

\label{lastpage}

\end{document}